\documentclass[letter]{aa} 

\usepackage{graphicx}
\usepackage{txfonts}
\newcommand{\doce}{\mbox{$^{12}$CO}}

\newcommand{\trece}{\mbox{$^{13}$CO}}

\newcommand{\jtd}{\mbox{$J$=3$-$2}}
\newcommand{\jdu}{\mbox{$J$=2$-$1}}

\newcommand{\kms}{\mbox{km\,s$^{-1}$}}

\newcommand{\ms}{\mbox{$M_{\mbox{\sun}}$}}

\newcommand{\lsim}{\raisebox{-.4ex}{$\stackrel{\sf <}{\scriptstyle\sf \sim}$}}
\newcommand{\gsim}{\raisebox{-.4ex}{$\stackrel{\sf >}{\scriptstyle\sf \sim}$}}

\newcommand{\secp}{\mbox{\rlap{.}$''$}}
\begin{document}

   \title{A second post-AGB
nebula that contains gas in rotation and in expansion: ALMA maps of IW Car}

   \author{V. Bujarrabal
          \inst{1}
          \and
          A. Castro-Carrizo\inst{2} 
          \and J. Alcolea\inst{3} \and H. Van
          Winckel\inst{4} \and C. S\'anchez Contreras\inst{5} \and
M.\ Santander-Garc\'{i}a\inst{3,6} }

   \institute{             Observatorio Astron\'omico Nacional (OAN-IGN),
              Apartado 112, E-28803 Alcal\'a de Henares, Spain\\
              \email{v.bujarrabal@oan.es}
\and 
 Institut de Radioastronomie Millim\'etrique, 300 rue de la Piscine,
 38406, Saint Martin d'H\`eres, France  
\and
             Observatorio Astron\'omico Nacional (OAN-IGN),
             C/ Alfonso XII, 3, E-28014 Madrid, Spain
\and
Instituut voor Sterrenkunde, K.U.Leuven, Celestijnenlaan 200B, 3001
Leuven, Belgium
\and
Centro de Astrobiolog\'{\i}a (CSIC-INTA), Ctra. M-108, km. 4,
E-28850 Torrej\'on de Ardoz, Madrid, Spain 
\and 
Instituto de Ciencia de Materiales de Madrid (CSIC). Calle
             Sor Juana In\'es de la Cruz 3, E-28049 Cantoblanco, Madrid,
             Spain 
   }

   \date{accepted 02 December 2016}

  \abstract
   {}
{We aim to study the presence of both rotation and expansion in
  post-AGB nebulae, in particular around IW Car, a binary post-AGB star
  that was suspected to be surrounded by a Keplerian disk.}
{We obtained high-quality ALMA observations of \doce\ and
  \trece\ \jtd\ lines in IW Car. The
  maps were analyzed by means of a simplified model of CO emission,
  based on those used for similar objects.}
{Our observations clearly show the presence of gas components in 
  rotation, in an equatorial disk, and expansion, which shows an
  hourglass-like structure with a symmetry axis perpendicular to the
  rotation plane and is probably formed of material extracted from the
  disk. Our modeling can reproduce the observations and shows moderate
  uncertainties.  The rotation velocity corresponds to a central
  stellar mass of approximately 1 \ms.  We also derive the total mass
  of the molecule-rich nebula, found to be of $\sim$ 4 10$^{-3}$
  \ms; the outflow is approximately eight times less massive
  than the disk. From the kinematical age of the outflow and the mass
  values derived for both components, we infer a (future) lifetime of
  the disk of approximately 5000--10000 yr.  }
   {}


   \maketitle
%

\section{Introduction}

Keplerian disks around post-AGB stars have been proven to be very
elusive. Gas in rotation has been directly observed in only two nebulae
to date, the Red Rectangle and AC Her (Bujarrabal et al.\ 2013b, 2015),
by means of interferometric mm-wave maps of CO lines.  The Red
Rectangle and AC Her belong to a class of binary post-AGB stars with
low-mass nebulae and with several independent lines of evidence of disks
(e.g., Van Winckel 2003; de Ruyter et al.\ 2006; Gezer et al.\ 2015;
Bujarrabal et al.\ 2013a).  They are characterized, in particular, by
spectral energy distributions (SEDs) with a NIR excess that indicates
hot dust close to the stellar system.  Single-dish observations of
\doce\ and \trece\ mm-wave emission in these post-AGB stars
systematically yielded characteristic line profiles, which are
strikingly similar to those of the Red Rectangle and AC Her and to
those expected to be emitted by relatively extended Keplerian disks
(Bujarrabal et al.\ 2013a).  A slowly expanding component was also
proposed to be present in this class of objects from those CO
data. ALMA maps of CO lines in the Red Rectangle indeed show a bipolar
low-velocity outflow (Bujarrabal et al.\ 2013b, 2016), very probably
formed of gas extracted from the disk and containing a mass
approximately ten times smaller.  Such a component was also confirmed
from CO maps of another of these NIR-excess post-AGBs, 89 Her
(Bujarrabal et al.\ 2007).  Rotation was not actually resolved in 89
Her, but a small disk could be confined to the prominent central
condensation. In this source (and probably in others observed in
single-dish), the contribution to the total emission of the outflow is
dominant and the outflow contains a mass at least comparable to that of
the compact disk. On the other hand, no sign of outflow was found in
the maps of AC Her, in which the expanding gas is probably very
diffuse.

The evolution of these objects is not well known (e.g., De Marco 2014)
and could be very different from that of high-mass (pre)planetary
nebulae. However, both kinds of sources share remarkable properties,
such as dominant axial symmetry, which has been proposed to be
associated with rotating disks (e.g., S\'anchez Contreras et al.\ 2002;
Soker 2001; Balick \& Franck 2002). Therefore, the study of our
objects, the only post-AGB ones in which disks are detected, could be
relevant to understand the formation of post-AGB nebulae in general.
Moreover, disks are observed in binary post-AGB stars that,
surprisingly, show orbits of insufficient size to accommodate an AGB
star (Gezer et al.\ 2015; Van Winckel et al.\ 2009). In the
best-studied nebula, the Red Rectangle (Bujarrabal et al.\ 2013b,
2016), the total angular momentum of the disk is not negligible and,
provided that all momentum comes from the binary system, would imply a
significant decrease of the distance between the stars.  Better
constraints on the structure and evolution of the disks
are therefore imperative to studying the orbital evolution of these
systems and their late evolution.

\begin{figure*}
  
   \hspace{.0cm}
   \includegraphics[width=18.cm]{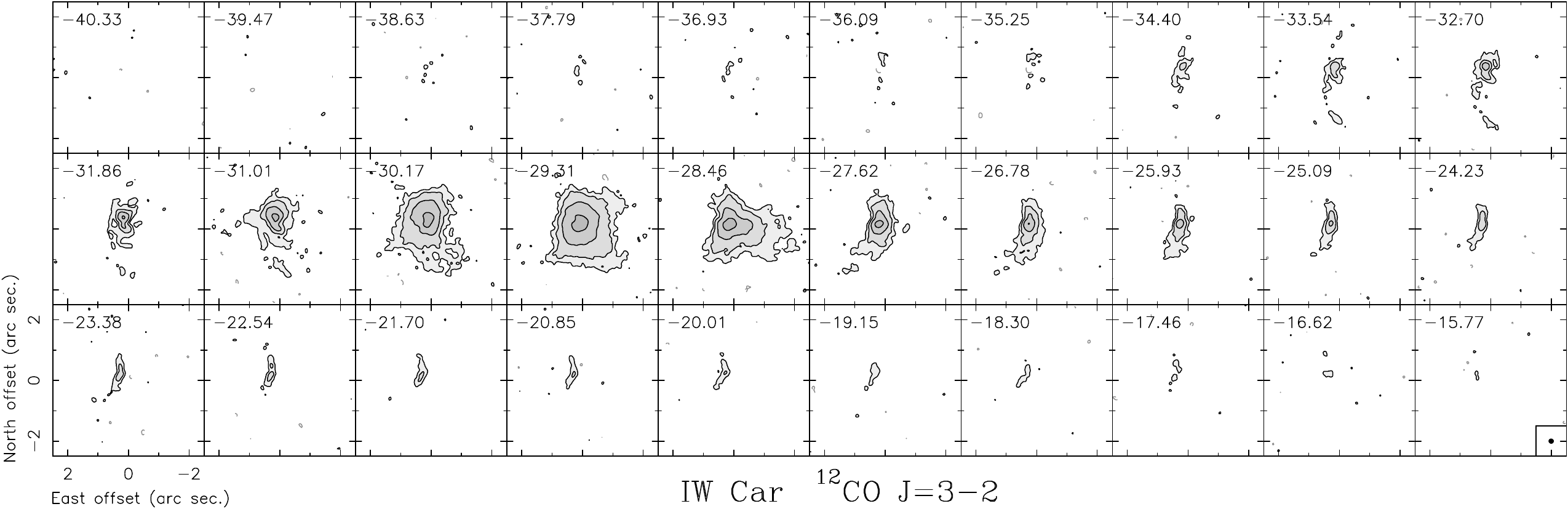}
      \caption{ALMA maps per velocity channel of \doce\ \jtd\ emission
        in IW Car. The continuum has been subtracted to better show the
        distribution of the weak line. The contour spacing is
        logarithmic: $\pm$5, 15, 45, and 135 mJy/beam (equivalent to
        $\pm$1.8, 5.1, 15.4, 46.2 K, in Rayleigh-Jeans equivalent
        temperature).
        The {\em LSR} velocities are indicated in each panel
        (upper-left corner) and the insert in the last panel shows the
        beam width.  }
         \label{}
   \end{figure*}

\begin{figure*}

     \hspace{-.0cm}
     \includegraphics[width=18.cm]{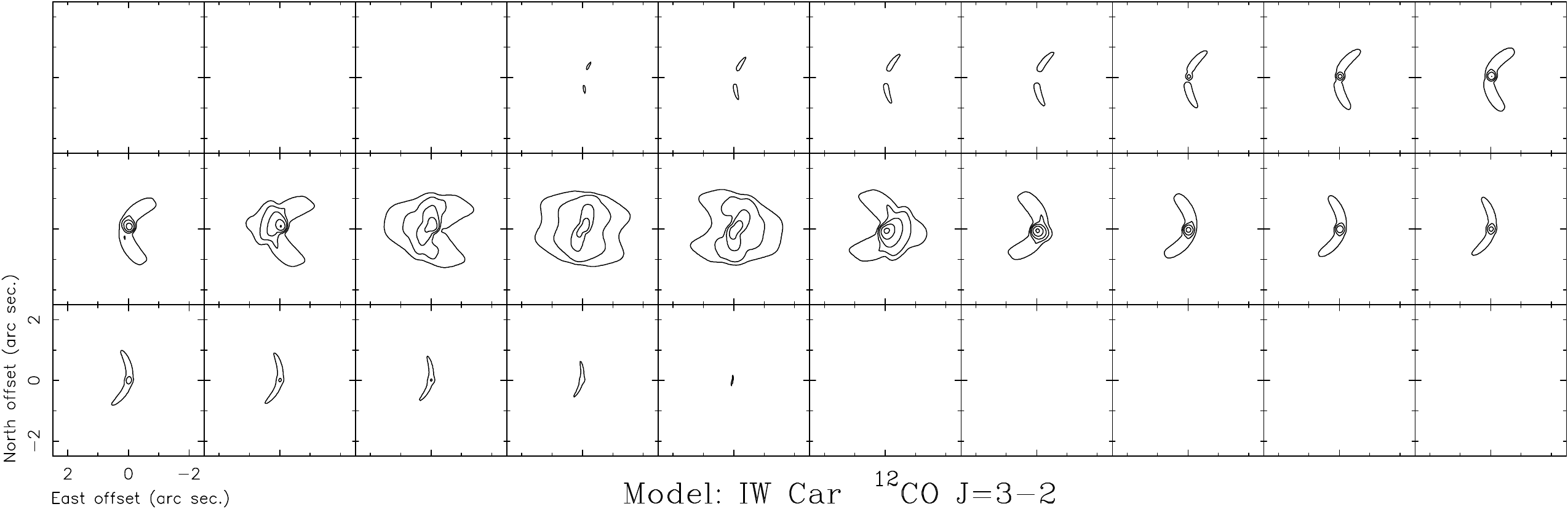}
      \caption{Predictions of our nebula model for \doce\ \jtd\ emission in
        IW Car (Sect.\ 3.1). To be compared with Fig.\ 1; all scales and contours
        are the same as in that figure.}
         \label{}
   \end{figure*}

We present ALMA maps of IW Car that clearly show both rotating and
expanding gas. A simplified model is compared with the observations,
allowing us to estimate the main nebular parameters. IW Car is an RV
Tau variable (Kiss et al.\ 2007) that belongs to the class of
NIR-excess post-AGB stars mentioned above (Giridhar et al.\ 1994; De
Ruyter et al.\ 2006; Bujarrabal et al.\ 2013a). A characteristic size
of the hot-dust component of $\sim$ 5 10$^{15}$ cm has been derived
from model fitting of the SED, although the nebular shape has not been
well studied.  IW Car is probably a double stellar system (with poorly
known orbital parameters) and shows a particularly high atmospheric
depletion.  Following those authors we adopt a distance of 1 kpc
for this source, but we stress that this value is uncertain.

\section{Observations}

We present maps of IW Car in the \doce\ and \trece\ \jtd\ lines,
$\lambda$ = 0.8 mm, obtained with ALMA, receiver band 7.  Four tracks
were consecutively observed on July 26, 2015, with an array
configuration of 42 antennas providing baselines between 12 and 1572
m. In total, $\sim$ 110 minutes of acquisitions were obtained on
source.  The data were first calibrated with the CASA software package.
The quasars J0522-3627 and J1058+0133 were used to calibrate the
bandpass, and J0842-6053 to calibrate gains. J0538-440 and J1107-448
were set as absolute flux references (with 0.91 Jy and 0.43 Jy at 345.8
GHz respectively). In addition, phase self-calibration was later
performed using the compact continuum source as a reference with the
GILDAS software package, which was also used in the following data
analysis. Image deconvolution was performed by using natural weighting,
leading to a resolution (HPBW) of approximately 0\secp 17$\times$0\secp
17.  The backends were set to achieve a spectral resolution of
approximately 0.21 \kms, which was degraded for our final maps because
of the poor S/N at high velocities, to a resolution of 0.85 \kms\ for
\doce\ \jtd\ (which shows emission at larger velocities) and of 0.66
\kms\ for \trece\ \jtd. By comparison with the APEX single-dish profile
(Bujarrabal et al.\ 2013b), we conclude that a small fraction of $\sim$
25\% of the flux has been filtered out in the maps of \doce\ \jtd. The
percentage of lost flux could be higher in the line wings, at $\pm$
3--4 \kms\ from the velocity center, but remains moderate throughout,
\lsim\ 50\%.  Dust continuum emission was detected, with total flux
$\sim$ 0.2 Jy, and barely resolved, with a tentative (deconvolved) size
$\sim$ 0\secp 1.

\begin{figure*}
  
   \hspace{-.cm}
     \includegraphics[width=18.2cm]{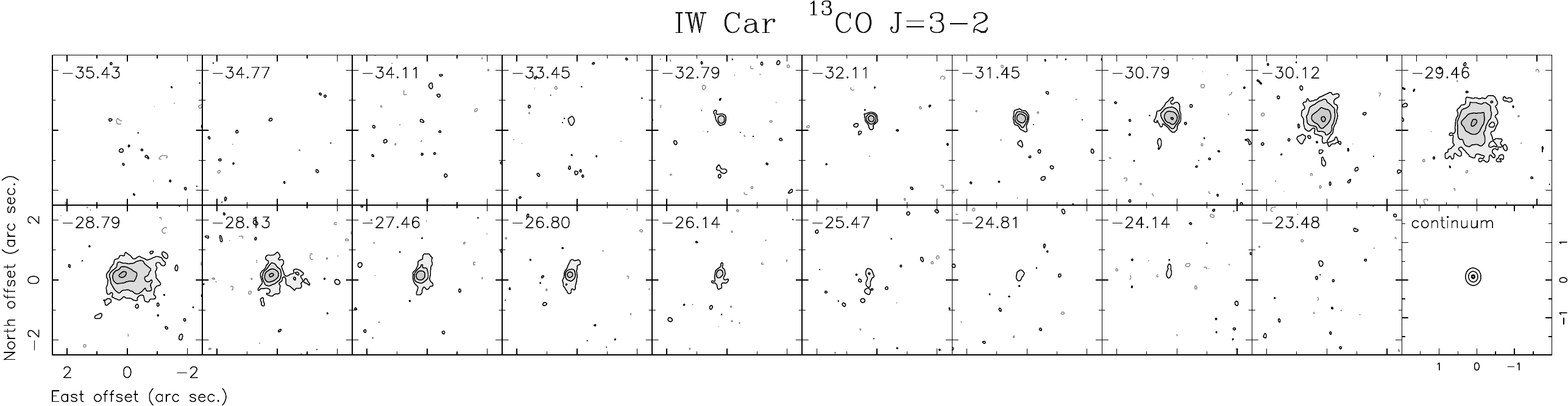}
      \caption{ALMA maps per velocity channel of \trece\ \jtd\ emission
        in IW Car. The continuum has also been subtracted in this
        figure. The contours are the same as in Fig.\ 1: $\pm$5, 15,
        45, and 135 mJy/beam (equivalent to $\pm$1.7, 5.3, 16.9, 47.8
        K).
        The {\em LSR} velocities are indicated in each panel (note that the
        velocity resolution is higher than in Fig.\ 1).  The last panel shows
        the continuum emission, contours: 5, 15, 45, and 135
        mJy/beam.}
         \label{}
   \end{figure*}

\section{Results}

In Fig.\ 1, we show our ALMA maps per velocity channel of the
\doce\ \jtd\ emission (in Fig.\ 2 we show the corresponding predictions
of our model, see below). We note the clear detection of an
hourglass-like structure, whose symmetry axis direction, projected in
the plane of the sky (position angle), is PA $\sim$ 75$^\circ$. The
structure is comparable to that found in the similar object 89 Her
(Sect.\ 1; but smaller in angular units). The strikingly similar
distributions of both structures strongly suggest that these lobes are
expanding, as confirmed by the model we describe below.  Fig.\ 3
shows our maps of \trece\ \jtd.  A hint of the lobes can be seen, but
their intensity is certainly lower than for \doce\ \jtd\ and the
central condensation largely dominates. A lower spectral resolution is
used for \doce\ \jtd\ because it shows the wide and weak emission from
the outflow, better seen in wider channels.

Rotation is also clearly detected in our maps of IW Car. Figs. 4 and 5
show position-velocity diagrams along the direction perpendicular
to the outflow axis, which corresponds to the equatorial plane,
at --15$^\circ$. The Keplerian pattern is very
obvious in both \doce\ and \trece\ diagrams. We recall that 89 Her
shows expanding lobes similar to those found here, but no rotation
was detected in it, although Bujarrabal et al.\ (2007) argued in favor
of the presence of rotation within the central unresolved clump.

Among the direct observational facts, we underline the particularly
high peak brightness found in the maps, of almost 100 K.  These high
values, very similar to the high brightness found in the Red Rectangle
(Bujarrabal et al.\ 2013b, 2016), and the comparable peak intensities
found for \doce\ and \trece\ \jtd\ suggest that the emission of the
central parts of the disk is optically thick (mostly for \doce) and
that the temperatures in IW Car should be high and similar to
those deduced for the Red Rectangle, \gsim\ 100 K in central regions. We
also point out the presence of blueshifted absorption of the relatively
intense emission of central regions by outer and cooler expanding gas
approaching us, as shown by the low negative contours that appear at
approximately --35 \kms\ in Figs.\ 1 and 4; this effect is expected in
expanding nebulae and is in fact also found in our modeling, see further
discussion in App.\ A.3.

\subsection{Simple modeling of our ALMA maps}

The S/N ratio and angular resolution of our data are moderate and,
moreover, only maps of \doce\ and \trece\ \jtd\ are available. We also
lack information on the nebula in general, including only single-dish
observations of \doce\ \jdu\ and \jtd. Under these conditions, very
detailed models, as those developed for the Red Rectangle, are not
sensible and we only perform simplified modeling.  See the predictions
of our model in Fig.\ 2 for \doce\ \jtd\ and a general description of
the model nebula in Fig.\ 6; the observed maps are reasonably
reproduced.  See further discussion and details on our modeling in
App.\ A.

In view of the similar outflows and disks found in IW Car and in 89 Her
and the Red Rectangle (Sect.\ 1), we base our modeling on the
properties we previously derived for those objects.
The width of the disk is not easy to determine from the data because of
the relatively low angular resolution, we adopt a
shape similar to that found in the Red Rectangle. In our best-fit
model, the expansion velocity is radial and its modulus is assumed to
vary with the distance to the equator and to the axis (see
App.\ A.1). The disk rotation is Keplerian with $V$(2\,10$^{15}$cm) =
2.5 \kms, which corresponds to a central stellar mass of $\sim$ 1 \ms;
this mass value is reasonable for a low-mass post-AGB double system,
though it is smaller than that found in the Red Rectangle ($\sim$ 1.7
\ms). The velocity value is uncertain by approximately $\pm$ 25\%
(also taking into account the uncertainty in the inclination, see
below). The uncertainty of the deduced central mass is 1$^{+0.6}_{-0.4}$ \ms. The axis
inclination with respect to the plane of the sky is poorly determined;
it cannot be either too large or too small in order 
to explain the easy detection of both expansion and rotation. We adopt
a value of $\sim$ 45$^\circ$ ($\pm$ 15$^\circ$).

   \begin{figure}
   \centering
   \includegraphics[width=7.3cm]{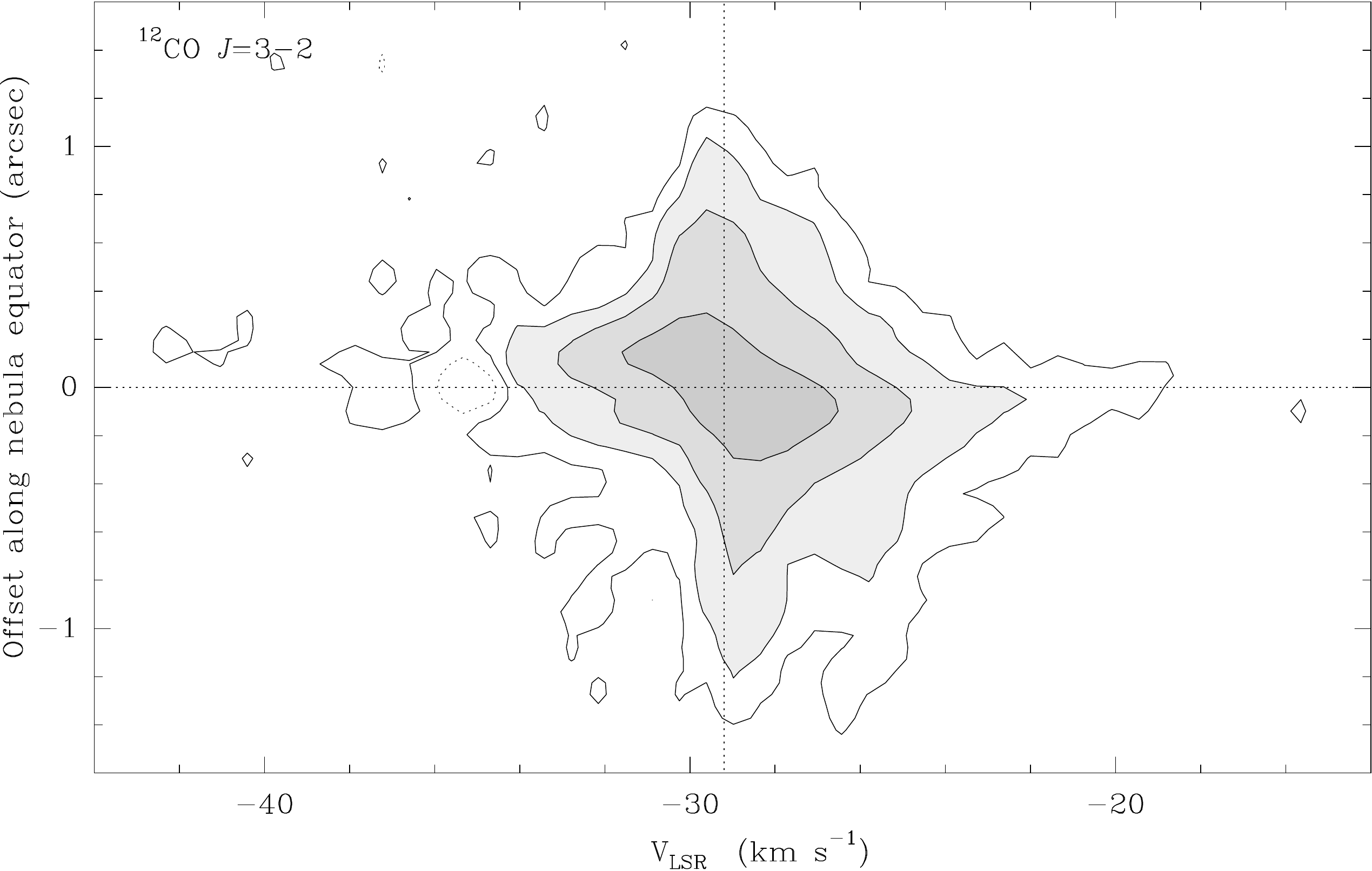}
      \caption{Position-velocity diagrams from our ALMA maps of
        \doce\ \jtd\ in IW Car along the direction P.A.\ =
        --15$^{\circ}$. Contours and the rest of the imaging parameters
        are the same as in the channel maps. The pointed lines show
        approximate centroids in velocity and position.}
         \label{}
   \end{figure}

   \begin{figure}
   \centering
   \includegraphics[width=7.3cm]{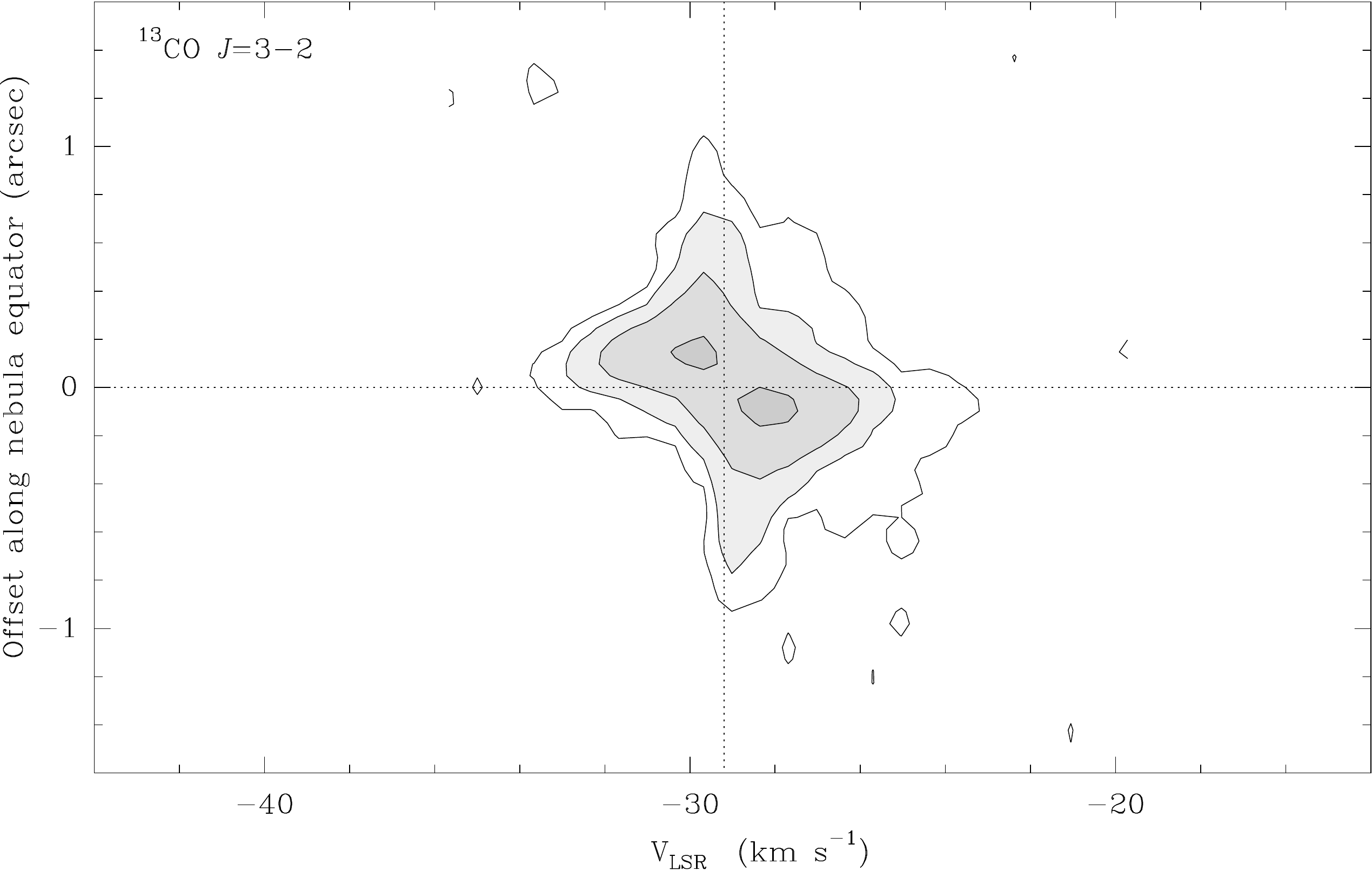}
      \caption{As in Fig.\ 1 but for 
        \trece\ \jtd.
      }
         \label{}
   \end{figure}
   
We assume LTE populations for the involved rotational levels. This is a
reasonable assumption for low-$J$ CO transitions in the dense material
expected in our sources, $n$ $>$ 10$^4$ cm, since their Einstein
coefficients are more than ten times smaller than the typical
collisional rates; see further discussion in Bujarrabal et al.\ (2013b,
2016).  The use of LTE may introduce some uncertainties (App.\ A.2),
but it simplifies significantly the calculations and provide an easier
interpretation of the fitting parameters.

We adopt a relative abundance $X$(\trece) $\sim$ 2 10$^{-5}$, to ease
the comparison with previous results on this object (Bujarrabal et
al.\ 2013a). In order to match the relatively low
\doce/\trece\ intensity ratio, we need a low abundance ratio, we deduce
$X$(\doce) $\sim$ 10$^{-4}$; similar low ratios are often found in
similar objects.  \trece\ \jtd\ is weaker than \doce\ \jtd, confirming
that the \trece\ emission is not very opaque. Our models yield values
of $\tau$(\trece\ \jtd) smaller than $\sim$ 0.3 in most of the disk and
significantly smaller in the outflow (see App.\ A.2).  The density and
temperature ($n$ and $T$) are assumed to depend solely on the distance
to the center and the equator with potential laws. See more details in
the Appendix; the density distribution is shown in Fig.\ 6.
The total mass derived from our fitting is 4
10$^{-3}$ \ms, very similar to the value found by Bujarrabal et
al.\ (2013a) from an analysis based on very
preliminary information on the properties of the nebula and only
\doce\ single-dish profiles. The mass of the outflow
is approximately 8 times smaller than that of the disk.

\section{Summary and conclusions} 

We present high-quality ALMA observations of \doce\ and
\trece\ \jtd\ line emission in IW Car.
IW Car belongs to a class of binary post-AGB stars that show properties
suggesting the presence of Keplerian disks around them (Sect.\ 1).
Maps per velocity channel, Figs.\ 1 and 3, show an hourglass-like
nebula in expansion, with an axis oriented along position angle PA
$\sim$ 75$^\circ$. This structure is very similar to that detected in a
similar object, 89 Her (Bujarrabal et al.\ 2007).  In the
position-velocity diagrams along the perpendicular direction, for PA =
--15$^\circ$, Figs.\ 4 and 5, we easily recognize the presence of a
disk in Keplerian rotation, probably in the equator of the
nebula. Similar rotating disks had been well identified before in only
two post-AGB nebulae, the Red Rectangle and AC Her, which also belong
to the same class of binary stars.  The expected presence of both
rotating and expanding gas had been well confirmed before in only the
best studied of these objects, the Red Rectangle, where the outflow was
shown to probably be extracted from the disk (Sect.\ 1).

   \begin{figure}
   \centering
   \includegraphics[width=7.7cm,angle=0]{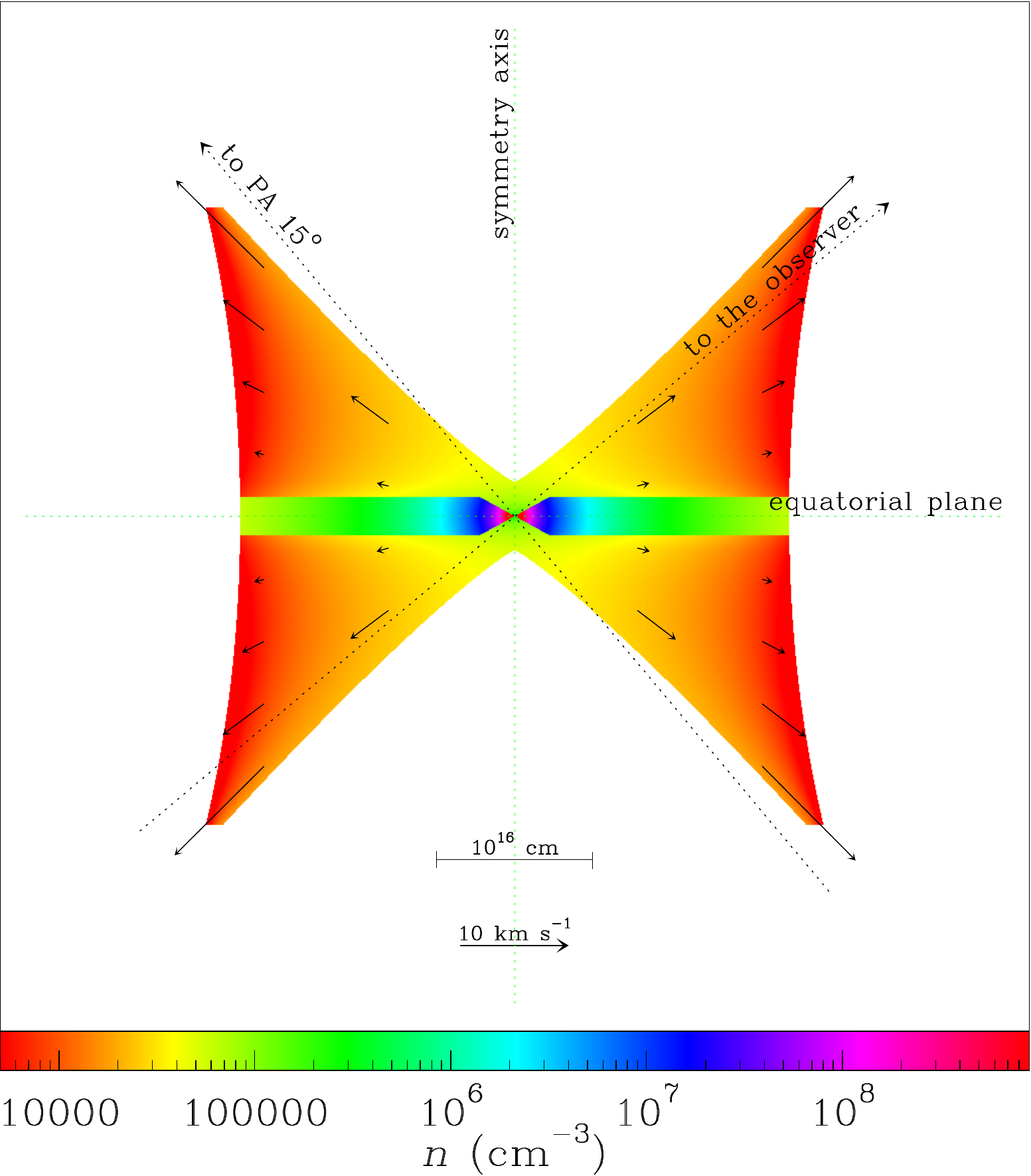}
      \caption{Structure and distributions of the velocity and density
        (in color) of our
        best-fit model for the disk and outflow. Only the velocity
          of the outflow is
        shown.  The inclination of the disk with respect to the line
        of sight is shown by the red-dashed line.}
         \label{}
   \end{figure}

Our modeling of the observations is simple but is able to reproduce
them and confirms our conclusions above (Sect.\ 3.1; Figs.\ 2 and
6). We find the Keplerian rotation to be compatible with a central
stellar mass of $\sim$ 1 \ms, reasonable for a post-AGB (double) star
of this kind. High temperatures of approximately 100 K were typically
found in the nebula. The total nebular mass is found to be $\sim$ 4
10$^{-3}$ \ms, with the mass of the outflow being approximately eight
times smaller than that of the disk. These mass values are $\sim$ 2.5
times lower than those found in the Red Rectangle, but the disk/outflow
mass ratios are very similar in both nebulae and the general structure
and dynamics are comparable.

In view of the similar properties we find in IW Car and the Red
Rectangle, we conclude that the expanding gas in IW Car also comes from
the disk.  From the mass values and outflow kinematics in IW Car, we
can derive the typical kinematical time required to form the outflow
($\sim$ 1000 yr) and the average mass-loss rate of the disk. We
therefore estimate the disk's future lifetime to be approximately 8000
yr. This value is similar to that found in the Red Rectangle, probably
longer than that estimated for 89 Her and shorter than for AC Her
(Bujarrabal et al.\ 2016). These lifetime values are comparable to or
longer than the expected post-AGB/preplanetary lifetimes of these
objects, and the disks could even survive in their (probable) planetary
nebula phase. We conclude that the coexistence of rotating equatorial
disks and gas in expansion extracted from it is probably a systematic
property of this class of post-AGB nebulae and that these are
long-living structures.

\begin{acknowledgements}
ALMA is a partnership of ESO (representing its member states), NSF
(USA) and NINS (Japan), together with NRC (Canada) and NSC and ASIAA
(Taiwan), in cooperation with the Republic of Chile. The Joint ALMA
Observatory is operated by ESO, AUI/NRAO and NAOJ. We made use of the
ALMA dataset ADS/JAO.ALMA\#2013.1.00338.S.  This work has been
supported by the Spanish MICINN, program CONSOLIDER INGENIO 2010 (grant
``ASTROMOL" CSD2009-00038), by the Spanish MINECO (grants AYA2012-32032
and FIS2012-32096), and by the European Research Council (ERC Grant
610256: NANOCOSMOS).
\end{acknowledgements}


\newpage

\appendix

\section{Further discussion on our simple modeling}

\subsection{Model description}

The determination of the structure and dynamics of the nebula
around IW Car is difficult because of the lack of information on this
source, in particular on the main properties of the nebula. For
that reason, we have chosen a simple modeling with few free
parameters. See a representation of out model nebula in Fig.\ 6 and our
model predictions in Fig.\ 2 and Figs.\ A.1 and A.2. The model nebula
is in some way a compromise between those we used to describe the
observations of 89 Her and the Red Rectangle (Bujarrabal et al.\ 2007,
2016).

In the rotating equatorial disk, we assume a purely Keplerian rotation
velocity in the central biconical region (Fig.\ 6), closer than $R_{\rm
  K}$ = 2 10$^{15}$ cm, with $V_{\rm rot}$($R_{\rm K}$ = 2.5 \kms). In
outer parts, we assume that there is also expansion at 3 \kms,
superposed to rotation, whose modulus in this case decreases inversely
proportional to the distance (following the law of angular momentum
conservation; the same was found for the outer part of the disk of the
Red Rectangle, which has been studied in much more detail). The density
of the disk is assumed to vary with the distance to the star following
a simple potential law, $n$ $\propto$ $r^{-2.7}$ with a value of 4
10$^5$ cm$^{-3}$ at half distance and 4 10$^6$ cm$^{-3}$ at the point
where the velocity changes to purely Keplerian, $R_{\rm K}$. The
temperature varies proportionally to $r^{-1.5}$, with a temperature of
$T$ = 200 K at $R_{\rm K}$.

In the hourglass-like outflowing component, we assume radial velocity,
with a modulus depending on the distances to the equator $h$ and to the
axis $p$ (varying proportionally to $h$ and linearly with $p_{\rm
  out}$--$p$, where $p_{\rm out}$ is the maximum value of $p$ for a
given value of $h$); see Fig.\ 6. This law is similar to that found for
the Red Rectangle. The temperature is assumed to vary again with
potential laws depending on the distance to the center: $T(r)$
$\propto$ $r^{-0.5}$, and a typical $T$(10$^{16}$cm) = 85 K. At
2 10$^{15}$ cm, high temperatures of
approximately 200 K are attained. The fitting is improved if we assume
a more complex variation for the outflow density, depending again on
$h$ and $p$; we adopted a law proportional to $h^{-0.5}$ and linear
with $p_{\rm out}$--$p$. Density values range typically between $\sim$
10$^4$ and 10$^5$ cm$^{-3}$, see Fig.\ 6.

\subsection{Uncertainties in the model parameters}

Some parameters describing the structure and dynamics show significant
uncertainties due to the lack of data on this nebula. The disk width is
not resolved, so its structure is uncertain. However, its diameter is
accurate within $\sim$ 20\%, given by the extent in the p-v
cuts. Variations of this parameter would yield inversely proportional
variations of the gas density, but not of its total mass, which must be
kept to explain the \trece\ measured intensities (see below).

The disk dynamics also show uncertainties. In particular, the point at
which the rotational velocity law changes is difficult to
determine. The outflow velocity field in the disk is also not easy to
describe, but the presence of two regimes is necessary, because
otherwise the predicted position-velocity cuts are significantly
different from the observational data. See an example in Fig.\ A.3 of
predictions in which a purely Keplerian field is assumed. The outflow
velocity modulus is constant and equal to 3 \kms, it probably
represents an order of magnitude of the departures from a Keplerian law
we must introduce to avoid these problems, more than a well defined
velocity field determined from model fitting. Due to the poor
information on this parameter and its meaning, we did not explore
possible complex laws, though we recognize that we probably followed an
overly simple description. However, the rotation velocity is better
constrained because we can identify it in the position-velocity
diagrams (see Sect.\ 3.1).

The main properties of the hourglass-like component are also difficult
to study in detail from the existing data.  We assumed radial
expansion, one of the simplest laws for the velocity. Radial expansion
is a simple definition and the expected case if launching takes place
along the force lines of a (locally) radial magnetic field. A linear
variation of the absolute value is in general expected if most
accelerations take place at the beginning of the process; in our case,
variation depending on the latitude and $p_{\rm out}$--$p$ (as for the
density, see A.1) leads to somewhat better predictions than dependence
on the distance to the center.  We can exclude some other simple cases,
such as constant velocity modulus, expansion in the axial direction,
and expansion parallel to the equator, for which the predictions do not
match the observations. However, other velocity laws, such as that
found for the outflowing gas in Red Rectangle by Bujarrabal et
al.\ (2016), which is more complex, would lead to results compatible
with the data (although involving a higher number of parameters). We
stress that the launching mechanism itself is obviously very uncertain
and its possible nature is a weak argument to support any velocity
field; a theoretical discussion on such processes is beyond the scope
of this letter. The total size, general shape, and overall velocity of
the outflowing gas are however relatively well constrained, since they
are given by the observed structure at several \kms\ from the central
velocity and the total velocity extent of the emission.

As mentioned in Sect.\ 3.1, the assumption of LTE level population
simplifies a lot our analysis. Nevertheless, we must keep in mind that
the {\em rotational temperatures} we use in our calculations may only
represent $J$-averaged excitation temperatures and not true kinetic
temperatures. Since high-$J$ levels can obviously be less populated in
reality than for LTE, these rotational temperatures are probably a
lower limit to the actual ones in the outflowing, relatively diffuse
gas.

Another consequence of the assumption of thermalized level populations
is that it is not possible to distinguish the effects of the density
and the relative CO abundances, $X$(\doce, \trece). Fortunately, $X$
seems relatively well constrained in the best studied objects,
particularly $X$(\trece), which is the basic parameter to determine the
total density and mass because of the lower opacity of
\trece\ lines. Our previous works always yielded $X$(\trece) ranging
between 10$^{-5}$ and 2 10$^{-5}$, both for post-AGB disks and for
young PNe
in general. We will adopt $X$(\trece) $\sim$ 2 10$^{-5}$ to ease the
comparison with previous results on this object (Bujarrabal et
al.\ 2013a). In order to match the relatively low
\doce/\trece\ intensity ratio, we need a relatively low abundance
ratio. We adopt $X$(\doce) $\sim$ 10$^{-4}$; similar low ratios are
often found in similar objects.  \trece\ \jtd\ is weaker than
\doce\ \jtd, confirming that the \trece\ emission is not very
opaque. Our model yields typical values of the optical depth
$\tau$(\doce\ \jtd) $\sim$ 3 in the disk (for the representative
velocities), while we typically find $\tau$(\doce\ \jtd) \lsim\ 0.2 in
the outflow emission.
For \trece\ \jtd, we find optical depths in the densest parts of the
disk $\tau$(\trece\ \jtd) $\sim$ 1. However, $\tau$(\trece\ \jtd)
$\sim$ 0.3 in most of the disk, and it is smaller than 0.1 in most of
the lines of sight intersecting the hourglass-like
component. Therefore, the determination of the density and mass of the
densest parts of the disk remains uncertain, but, except for this case,
the mass determination is relatively accurate, provided that we
reproduce the total \trece\ intensity. The value of the total mass is
mainly affected by the assumed $X$(\trece) value and depends slightly
on other parameters such as the geometry and kinematics.

\subsection{Comparison of the predictions with the observational data}

Fig.\ 2 shows the observational and synthetic maps per velocity channel
and Figs.\ A.1 and A.2 show the predicted position-velocity diagrams
along the rotating disk. The model reproduces most of the observed
features, but some problems are still present. The major one concerns
the faint extended emission, mostly at moderate velocities further than
approximately 2 \kms\ from the central velocity. The predicted features
are always somewhat wider. This is very probably due to the significant
loss of flux in the interferometric observation (Sect. 2).  This
relatively extended and weak emission is expected to be particularly
affected by the significant instrumental flux loss, which means that we
can expect noisier data and a significant overestimate of the
predictions.  Because of the steep logarithmic scale we use, changes at
a level of approximately 1/30 of the peak brightness (our first
contour) are very clearly shown in the figures, even if their effect on
the total emission is moderate. This issue also appears in the
synthetic position-velocity diagrams, at intermediate velocities and at
$\sim$ $\pm$ 0\secp 5 from the center, which show an overly extended
faint emission. We have tried to minimize
this effect in our modeling, but it cannot disappear without
significantly affecting other predictions, like the total velocity
extent, which become incompatible with observations.

We also note that the observed emission extends significantly more at
less negative {\em LSR} velocities; an asymmetry that we do not try to
introduce in our model, which assumes symmetry with respect to the
equator and axis. Obviously, the actual nebula does not exactly show
such symmetries, which is a common issue when modeling
post-AGB nebulae.

Another effect of the presence of expanding gas is the weak
absorption that appears at approximately --35 \kms. The absorption is
noticeable in the different shapes observed between relatively
redshifted and blueshifted features (Figs.\ 1 and 2) and in the
different blue/red sides of the p-v diagram (Figs.\ 4, A.1), with
blueshifted emission systematically weaker. This phenomenon depends on
subtle geometrical and velocity coincidences and is very difficult to
model. Our code predicts absorption features similar to those observed
but, because of the uncertain treatment of this effect, we will not try
to fit it.

   \begin{figure}[h]
   \centering
   \includegraphics[width=7.3cm]{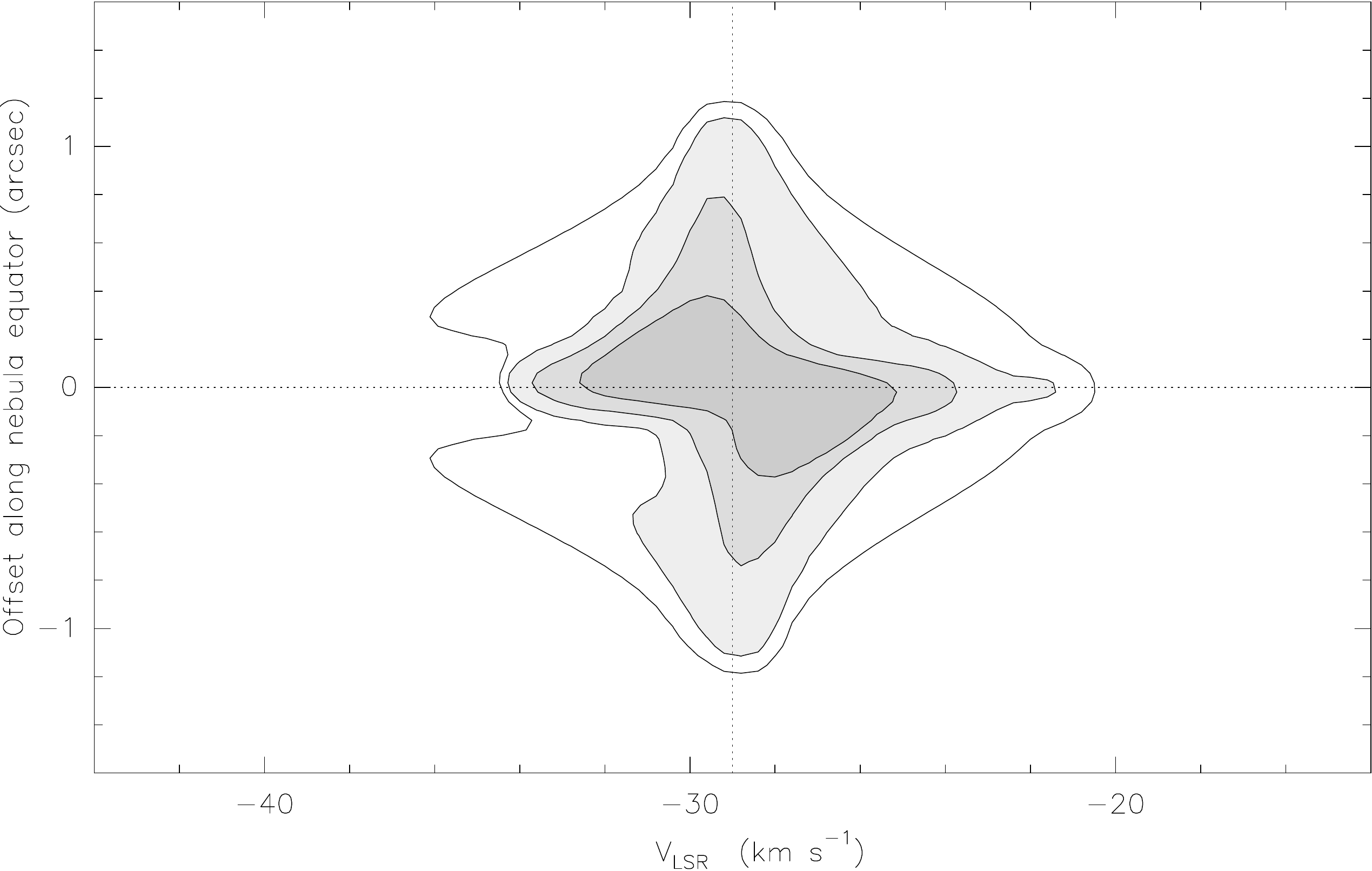}
      \caption{Position-velocity diagrams predicted from our simple
        model for the 
        \doce\ \jtd\ emission in IW Car along the equatorial
        direction, to be compared with Fig.\ 4; contours and the rest
        of the imaging parameters 
        are the same as in that figure.}
         \label{}
   \end{figure}

   \begin{figure}[h]
   \centering
   \includegraphics[width=7.3cm]{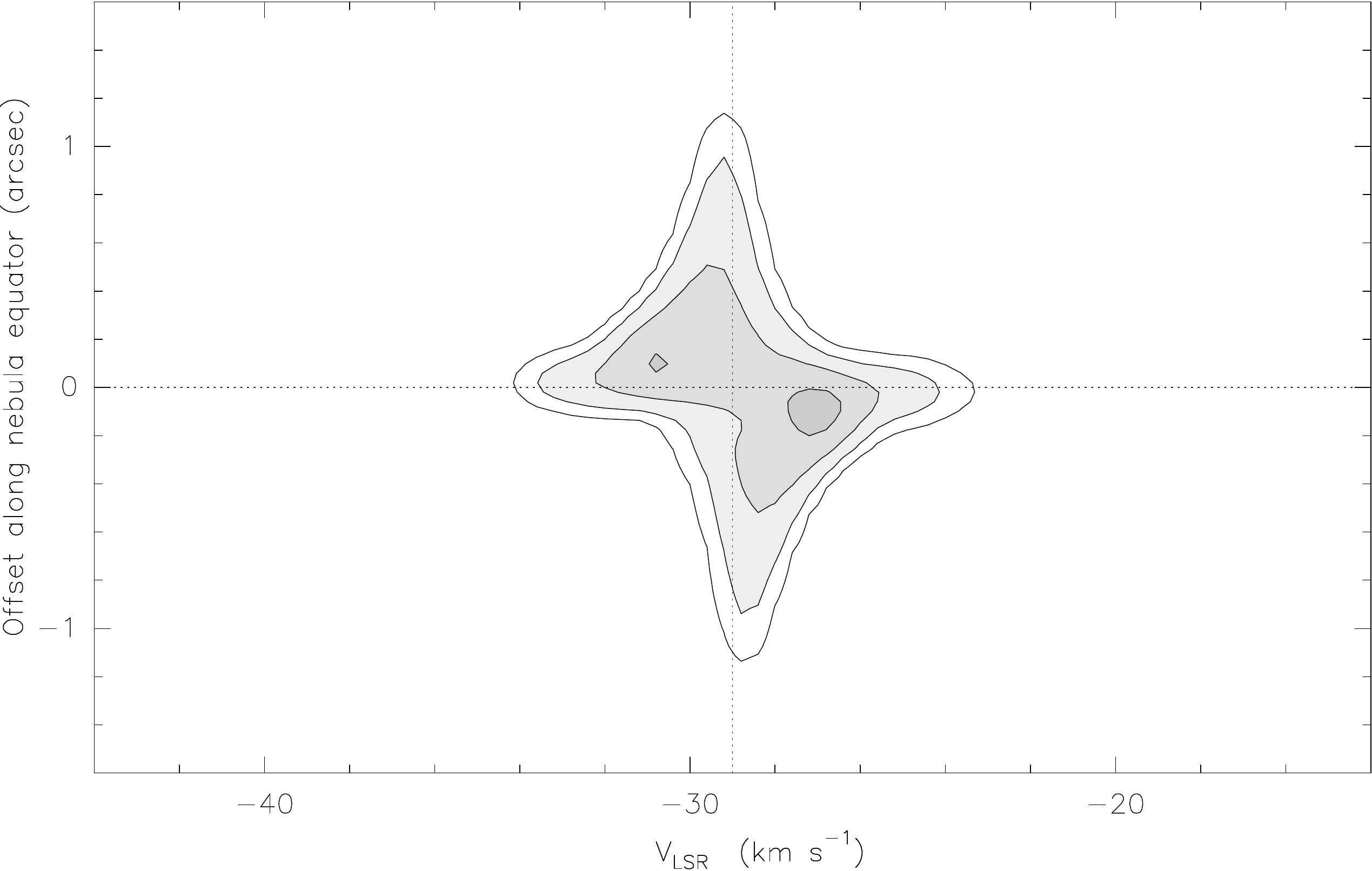}
      \caption{As for Fig.\ A.1 but for 
        \trece\ \jtd. Synthetic position-velocity diagram along the
        equatorial direction, to be compared with Fig.\ 5.
      }
         \label{}
   \end{figure}

   \begin{figure}[h]
   \centering
   \includegraphics[width=7.3cm]{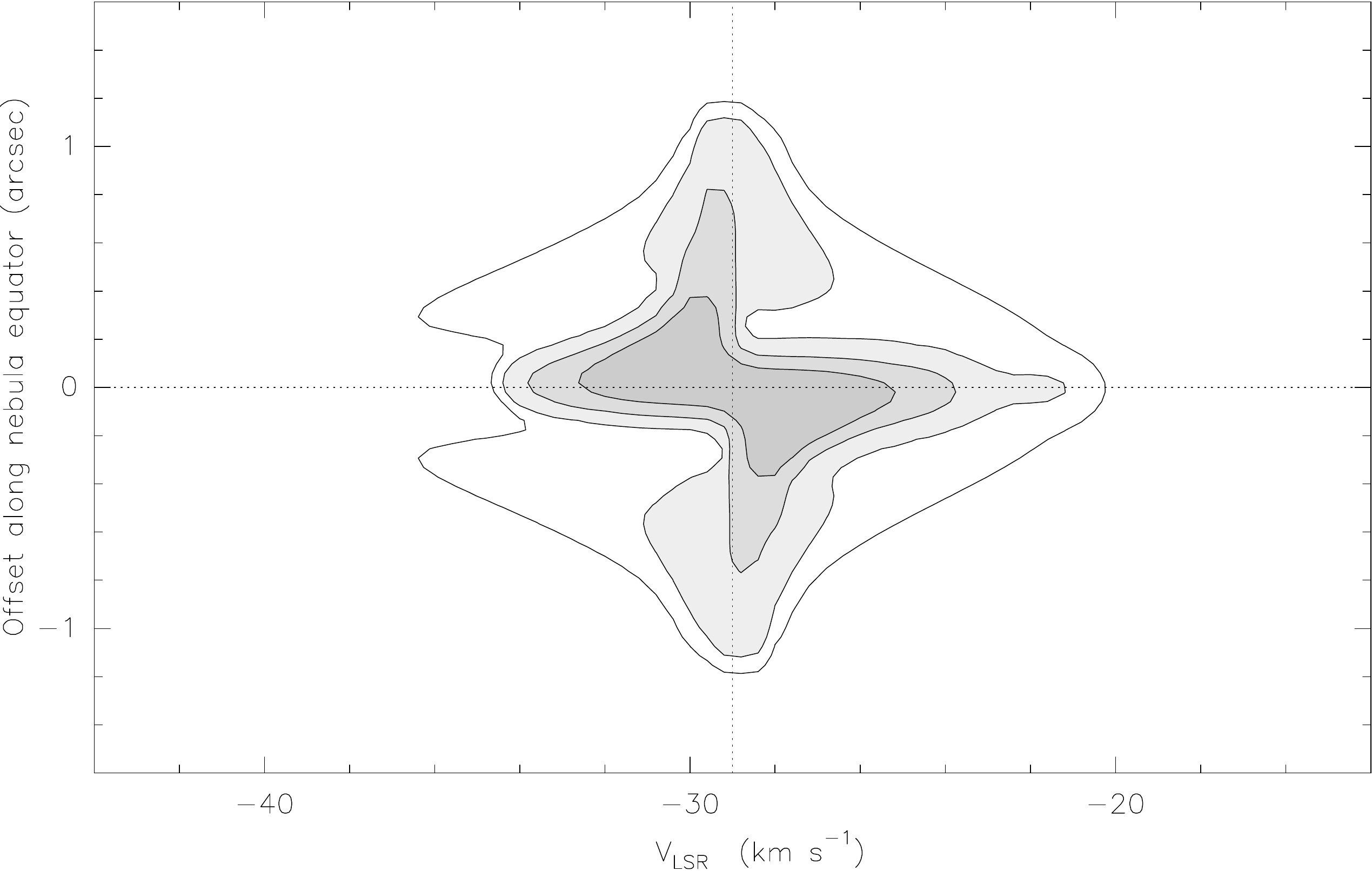}
      \caption{As for Fig.\ A.1 but assuming that the outer disk shows
        no expansion velocity. The rest of the model parameters are
        kept the same. The predicted position-velocity field is
        significantly different from the observations (Fig.\ 4).
      }
         \label{}
   \end{figure}

\end{document}